\documentclass[amsmath,amssymb,12pt]{article}
\usepackage{jheppub}
\usepackage[utf8]{inputenc}
\usepackage{amsfonts}
\usepackage{amsthm}
\usepackage[cal=dutchcal]{mathalfa}
\usepackage{graphicx}
\usepackage{slashed}
\usepackage{xcolor}
\usepackage{float}
\usepackage{slashed}
\usepackage{bm}
\usepackage[T1]{fontenc}
\usepackage{lmodern}

\usepackage{cancel}
\usepackage[normalem]{ulem}
\usepackage{cancel}
\newcommand{\comment}[1]{}

\usepackage{cleveref}

\title{On type-II Spacetimes and the Double Copy for Fluids Metrics}

\author{Cynthia Keeler$^{\dagger}$ and Nikhil Monga$^{\ddagger}$}
\date{April 2024}
\emailAdd{$^{\dagger}$keelerc@asu.edu \& $^{\ddagger}$nmonga.academic@gmail.com}
\affiliation{$^{\dagger}$Physics Department,
	Arizona State University, Tempe, AZ 85287, USA}
\date{May 2019}
\abstract{In our previous paper \cite{keeler2020navierstokes} we discussed type-D and type-N fluid-dual spacetimes and provided their associated single copies in the context of the Weyl double copy. In this work we extend our analysis to more general fluids thereby requiring the application of the double copy picture to type-II space-times. By combining our previous type-D and type-N fluids via their associated stream functions we demonstrate an example of a viable type-II double copy. Further we use an explicitly perturbative approach in the near horizon expansion to generalize the type II double copy for the fluid-dual space-times. We show a Maxwell spinor ansatz containing a heterogeneous bi-spinor component is necessary to provide a viable type-II double copy at the lowest order.
}

\begin{document}

\maketitle

\section{Introduction}	

The Parke-Taylor formula of \cite{Parke:1986gb} succinctly demonstrates how n-gluon tree amplitudes, if they are MHV simplify significantly.  This seminal work, \cite{Parke:1986gb}, has translated into a broader understanding of how closed string amplitudes can be written as squares of open string amplitudes \cite{Kawai:1985xq,Bern:1999bx,Bern:1999ji,Bjerrum-Bohr:2003hzh,Bjerrum-Bohr:2003utn}. A culmination of these insights results in the double copy paradigm \cite{Bern:2008qj,Bern:2010ue,Bern:2010yg,Cachazo:2013gna}. Fundamentally the double-copy picture posits that tree level gravitational amplitudes can be ``double-copies'' of Yang-Mills amplitudes.

Recent formulations of the classical double copy make more explicit this classical/tree-level mapping between spin-2 and spin-1 fields. Work by \cite{Monteiro:2014cda} demonstrates that space-times which admit a Kerr-Schild like expansion, i.e. when written as, 
\begin{equation}
	g_{\mu\nu} = \eta_{\mu\nu} + \phi k_\mu \tilde{k}_{\nu},
\end{equation}
allow a mapping between the null vector fields $k_{\mu}$ and gauge fields $A_{\mu}^a$. This allows us to identify Yang-Mills fields as,
\begin{equation}
	A_{\mu}^a \sim k_{\mu}c^a.
\end{equation}

The Weyl double copy [14] provides a second formulation, using spinor notation to identify the Weyl spinor with the field strength spinors via,

\begin{equation}
	C_{ABCD} = \frac{1}{S}\,f_{(AB}f_{CD)}.
\end{equation} 

The Weyl double copy was extended to arbitrary Petrov types by \cite{Chacon:2021wbr} by utilizing a twistor space representation (see \cite{White:2020sfn,Elor:2020nqe,Chacon:2021lox,Luna:2022dxo} for more). More recently \cite{Chawla:2022ogv} construct examples of the Weyl double copy in general dimensions and provide novel examples in five dimensions. See \cite{Kosower:2022yvp,Adamo:2022dcm} for recent reviews.

We proceed from our previous work in \cite{keeler2020navierstokes} where we demonstrate how the Weyl double copy of \cite{Monteiro:2018xev} with the fluid-gravity duality of \cite{Bredberg:2011jq} can be utilized to map solutions of the Navier-Stokes equations to the Maxwell's equations. We obtained a simple mapping between components of the fluid and their associated single and double copy fields.

On the gravity side, we found that irrotational potential fluid flows are associated with type-N spacetimes, while constant vorticity fluids are associated with type-D spacetimes. With respect to their associated single copy gauge fields, the type-D, constant vorticity fluids allowed for a simple mapping between the fluid velocity onto gauge fields, or equivalently allowing the vorticity of the fluid to be proportional to the magnetic field of the single-copy gauge field. The scalar potential associated with the potential flow of the type-N spacetime was mapped onto the zeroth copy scalar field.

However, in \cite{keeler2020navierstokes} in order to map the single copy fields onto fluids it was necessary to reduce the type-II fluid metric onto either type-N or type-D metrics. In this paper we extend our previous analysis and demonstrate that type-II fluid metrics are amenable to the double copy procedure. This stream function allows us to specify all two dimensional incompressible fluids.

Although we work with a specific metric, which we write as a near-horizon expansion to Rindler space, throughout this paper we have attempted to write expressions in terms of generic geometric quantities --- specifically Weyl scalars $\Psi_2$ and $\Psi_4$. We do this as the restrictions we find in solving the Maxwell and scalar equations should be translatable to other space-times that can similarly be expressed as expansions near Rindler horizons. 

The structure of this paper is as follows: in \cref{methods} we review the near-horizon expansion in the context of the fluid-gravity duality, highlighting our previous work \cite{keeler2020navierstokes} and restating its results in terms of a fluids stream function. In \cref{section_simpletype2} we provide an example of a simple type-II fluid-dual metric by combining our previous type-D and type-N metrics.  In \cref{genfluidsdc} we consider a more general form of single and zeroth copy fields for type-II fluid-dual spacetimes.  In \cref{onvorticityeq} we relate the fluid vorticity equation and stream function to the Weyl scalars, and also explore the single copies of spacetimes dual to some exact Navier-Stokes solutions.


\section{The Fluids Metric \& the Stream Function}\label{methods}

In this section we summarize the methodology used. We first describe the near horizon limit of the fluid-dual metric as stated in \cite{Bredberg:2011jq} and used by us previously in \cite{keeler2020navierstokes}, then we identify the Weyl scalars thus obtained and map them onto a scalar function for the fluid, the stream function, and finally we discuss the conditiions necessary on these functions so as to solve all the necessary equations of motion. 
\subsection{The Fluids Metric in the Near Horizon Limit And Steps to a Type-II Double Copy} 
In the near horizon expansion the fluid-dual metric takes the form, 

\begin{equation}
\begin{split}
		\label{lambdaexpansion}
		d\hat{s}^2 =&-\frac{ r  }{\lambda} d \tau ^2 \\ & + \left[2 d\tau dr + dx_i dx^i -2 (1-r) v_i d x^i d\tau +{(1-r)}(v^2+2P) d\tau^2  \right]
		\\
		& +\lambda\big[{(1-r)}{v_i v_j }dx^i dx^j -2{v_i}dx^i dr   + (v^2+2P)d\tau dr
		\\
		&+
		(r-1)[-(r+1) \partial^2 v_i + (v^2+2P)2v_i +4 \partial_i P]dx d\tau\big] + \mathcal{O}(\lambda^2).
\end{split}
\end{equation}
Einstein's equations for this metric correspond to the Navier-Stokes equations for an incompressible fluid, specifically, we see, 
\begin{equation}
	\begin{split}
		G_{00} &= 0 \implies \partial_i v_i = 0, \\
		G_{0i} &= 0 \implies \partial_\tau v_i-\eta\partial^2 v_i+\partial_i P+v^j\partial_j v_i=0,
	\end{split}    
\end{equation}
and the Weyl scalars for this metric are, 
\begin{equation}
	\Psi_2 = \frac{-i}{4} (\partial_x v_y - \partial_y v_x) \qquad \Psi_4 = \frac{-i \,}{2} (\partial_x v_y + \partial_y v_x - i(\partial_x v_x - \partial_y v_y)).
\end{equation}
For fluids in two dimensions, we can characterize flows by a single scalar whose curl provides the velocity of the fluid. This scalar is referred to as the stream function. That is we have, 
\begin{equation}
	v_i  = \epsilon_{ij} \partial_j \chi.
\end{equation}
Incompressibility follows trivially from the above relation since the divergence of a curl vanishes. Taking a curl of the above expression, we can see that the Laplacian of the stream function is related to the vorticity of the fluid, 
\begin{equation}\label{psi2vorticity}
\partial^2 \chi = -\omega(x,y).
\end{equation}
Using the above, the Weyl scalars can be expressed directly in terms of this stream function.\footnote{Note that these quantities are more clearly expressed in terms of holomorphic and anti-holomorphic functions, and thus for the remainder of this paper we will be writing the stream function in terms of complex coordinates.}
\begin{equation}\label{psisol}
	\Psi_2 = i \partial_z \partial_{\bar{z}} \chi \qquad \Psi_4 = 2 i  \partial_{\bar{z}}^2 \chi.
\end{equation}
This identification of the Weyl scalars in terms of the stream function will have utility as we probe the existence of type II fluid-dual metrics that satisfy the double copy. 

\subsubsection{Revisiting Type D and Type N Double Copies}

We begin by revisiting the type-D and type-N double copies in \cite{keeler2020navierstokes} written in terms of the stream function $\chi$. We require the following conditions to hold for type-D and type-N space-times respectively:

\begin{enumerate}
	\item \textbf{For type-D space-times} the Weyl scalar $\Psi_4$ must vanish (see \cite{Stephani:2003tm} for a comprehensive overview), that is, from eqn. \eqref{psisol} we require that the stream function $\chi_D$ be such that, 
	\begin{equation}
		\psi_4 = 2i\partial_{\bar{z}}^2 \chi_D = 0. 
	\end{equation}  
Along with the above, requiring the type-D fluid velocity to be real gives us the following form for the type-D stream function as, 
\begin{equation}\label{psi2val}
	\chi_D = A(\tau)z+\bar{A}(\tau)\bar{z} + B(t)z\bar{z}.
\end{equation}
Such a stream function is associated with a spatially constant Weyl scalar $\Psi_2$, more precisely from \cref{psi2vorticity,psisol} we find,
\begin{equation}\label{psi2}
	\Psi_2 = - i \frac{\omega }{2} = i B(\tau). 
\end{equation}
	\item \textbf{For type-N space-times} we can  identify the stream function such that the Weyl scalar $\Psi_2$ vanishes giving us a type-N fluid. Therefore we find,  
\begin{equation}
	\Psi_2 = i \partial_{z} \partial_{\bar{z}} \chi_N = 0.
\end{equation}

A stream function $\chi_N$ that is either a sum of holomorphic and anti-holomorphic pieces easily satisfies this contraint. Such a stream function can be identified with a fluid potential associated with a vanishing Laplacian. Such fluid flows are commonly referred to as potential flows. The associated stream function can be written as 
\begin{equation}\label{psi4}
	\chi_N = -i f(z,\tau) + i \bar{f}(\bar{z},\tau).
\end{equation}
\end{enumerate}

In the next section we first describe a simple example of a type-II double copy obtained by simply combining the type-D and type-N fluid stream functions above in \eqref{psi2} and \eqref{psi4}. Subsequently we examine more general type-II double copy fields and discuss the constraints associated with satisfying the associated Maxwell and scalar  field equations of motion. 

\section{A Simple Type II Fluids Double Copy}\label{section_simpletype2}
The Weyl spinor associated with a type-II spacetime can be written as\footnote{Note however, a Weyl spinor with other non-vanishing components may also be type-II, however it can be rotated in to the provided form via a tetrad rotation.}
	
\begin{equation}\label{cabcd}
	C_{ABCD} = 6 \Psi_2\, o_{(A} \iota_{B}o_{C} \iota_{D)} + \Psi_4\, o_{A} o_{B}o_{C}o_{D}.
\end{equation}

In the context of the double copy this Weyl spinor is mapped to a zeroth copy $S$ and a mixed single copy, with Maxwell spinors $f^{(1)}_{AB}$ and $f^{(2)}_{AB}$. In contrast to the type-D case, the associated Maxwell spinors need not be identical,

\begin{equation}\label{type2dc}
	\frac{1}{S} f^{(1)}_{(AB}f^{(2)}_{CD)} = C_{ABCD}.
\end{equation}

The simplest type-II double copy can be constructed by combining stream functions for type-D and type-N fluids via,
\begin{equation}
	\chi_{II} = \chi_D+\chi_N.
\end{equation}

A simple stream function that provides us with a type-II double copy of the form as written in \eqref{type2dc} will  correspond to a velocity at most linear in coordinates and a spatially invariant vorticity. The stream function for such a fluid is

\begin{equation}\label{chitypeii}
	\chi(\tau,z,\bar{z}) = A(\tau)z+\bar{A}(\tau)\bar{z} + B(\tau)z\bar{z} + i \bar{C}(\tau)\bar{z}^2  - i C(\tau) z^2.
\end{equation}
Utilizing the mapping between the stream function and Weyl scalars as provided in \eqref{psisol}, we propose the following Weyl scalars for this type-II spacetime,   
  
\begin{equation}\label{simpletypeiiweylscalars}
	\Psi_2 = i B(\tau) \qquad \Psi_4 = -4 \bar{C}(\tau).
\end{equation}


These Weyl scalars do not provide the most general single copy fields. However, the single copy fields do satisfy their respective spin-1 Maxwell and spin-0 wave equation on the Rindler-like background.  

The form of our Maxwell spinors that follows from \eqref{type2dc} is

\begin{equation}\label{fiAB0type2}
	f_{AB}^{(i)} = P^{(i)}o_{(A}\iota_{B)} + Q^{(i)}(\tau), o_{(A}o_{B)}.
\end{equation}
Here in order to satisfy equations of motion $P^{(1)}$ and $P^{(2)}$ must be constant, while $Q^{(1)}$ and $Q^{(2)}$ can only depend on time. Additionally Q, B, $\bar{C}$ and the constant P must satisfy,
\begin{equation}\label{q1abcond}
	Q^{(1)}(\tau)^2 =i\,\frac{2}{3} \frac{\bar{C}(\tau)}{B(\tau)}{P^{(2)^2}} \quad \textrm{and} \quad	P^{(1)} Q^{(2)}+P^{(2)} Q^{(1)} = 0.
\end{equation}

The associated zeroth copy that solves the scalar wave equation is, 

\begin{equation}\label{stypeii}
	S_{II} = -i \frac{P^{(1)}P^{(2)}}{\bar{C}(\tau)}  \implies \Box S_{II} = 0.
\end{equation}

The Maxwell and scalar wave equations we solve live on a Rindler-like background metric. This metric can be obtained from the full metric in equation \eqref{lambdaexpansion} by setting the fluid velocity and pressures to vanish:

 
\begin{equation}
	\label{metric_background_app}
	d\hat{s}^2 =-\frac{ r  }{\lambda} d \tau ^2  + 2 d\tau dr + dx_i dx^i.
\end{equation}

Taken together \cref{chitypeii,simpletypeiiweylscalars,fiAB0type2,q1abcond,stypeii} provide us with a viable type-II double copy of the form \eqref{type2dc} which satisfies the scalar wave and Maxwell equations on the background metric above in \cref{metric_background_app}.

In the next section we consider more general type-II solutions and examine the constraints that arise while trying to satisfy the equations of motion for the associated single and zeroth copies.

\section{Generalizing The Type II Double Copy}\label{genfluidsdc}

In the previous section we examined how a type-II fluid double copy was achievable, however for a relatively simple fluid velocity profile, one with a velocity at most linear in its spatial coordinates and a vorticity that is spatially invariant.\\

In this section we attempt a more general double copy and identify the constraints that present themselves while extending the double copy picture for the fluds metric more complex fluid flows. To do so, in the next two subsections we list out the general functional forms of these Maxwell spinors and as well as the scalar that satisfy the Maxwell and scalar wave equations on the background. 

\subsection{General Solution for the Scalar Equation on the Fluids Background}

For the scalar function $S$, we identify a perturbative expansion in $\lambda$ of the type, 
\begin{equation}
	S \equiv S^{(0)}  +  \lambda S^{(1)} + \lambda^{2} S^{(2)} + O(\lambda^3).
\end{equation}
We are intrested in identifying a general functional form of this function such that the scalar equation $\Box S=0$ on the background metric as stated in \cref{metric_background_app} above are satisfied, plugging the above ansatz into $\Box S=0$ we find, 
\begin{align}
	O(\lambda^{-1}): &\,\, \partial_{r}(r\, \partial_r S^{(0)}) = 0, \label{lambdaminus1S}\\
	O(\lambda^{0}): &\,\,  \partial_{r}(r\, \partial_r S^{(1)}) + 2\partial_{\tau} \partial_{r} S^{(0)} + \partial_x^2 S^{(0)} + \partial_y^2 S^{(0)} = 0, \label{lambdazeroS}\\
	O(\lambda^{1}): &\,\,  \partial_{r}(r\, \partial_r S^{(2)}) + 2\partial_{\tau} \partial_{r} S^{(1)} + \partial_x^2 S^{(1)} + \partial_y^2 S^{(1)} = 0. \label{lambdaone}
\end{align}
At higher orders a similar pattern as the one noted in \cref{lambdaone} continues, i.e., we have, 

\begin{equation}
	O(\lambda^{i}): \,\,  \partial_{r}(r\, \partial_r S^{(i+1)}) + 2\partial_{\tau} \partial_{r} S^{(i)} + \partial_x^2 S^{(i)} + \partial_y^2 S^{(i)} = 0 \qquad \forall \,\, i>0. 
\end{equation} 
At the lowest order, $O(\lambda^{-1})$, from eqn. \cref{lambdaminus1S} we have, 
\begin{equation}\label{scalarsol0}
	\partial_{r}(r\, \partial_r S^{(0)}(\tau,r,x,y)) = 0 \implies S^{(0)} = S^{(0)}_a (\tau,x,y) + \log(r) S^{(0)}_b(\tau,x,y).
\end{equation}
Plugging this lowest order solution for the Maxwell scalar we have at the next higher order at $O(\lambda)$,
\begin{equation}
	\partial^2 ( S^{(0)}_a (\tau,x,y) + \log(r) S^{(0)}_b(\tau,x,y)) + \frac{2}{r} \partial_{\tau}S^{(0)}_b(\tau,x,y)_+ \partial_r(r\partial_r S^{(1)}(\tau,r,x,y))=0.
\end{equation}
The above expression can be integrated in r to identify the r-dependence for the next higher order piece in the scalar function $S^{(1)}$. We have, 
\begin{equation}\label{scalarsol1}
	\begin{split}
		S^{(1)}(\tau,r,x,y) &= \log(r) S^{(1)}_a(\tau,x,y)+S^{(1)}_b(\tau,x,y) - r \partial^2 S^{(0)}_a(\tau,x,y) \\&\hspace{1cm}- r(\log(r)-2) \partial^2 S^{(0)}_b(\tau,x,y) - \log(r)^2 \partial_{\tau} S^{(0)}_b(\tau,x,y).
	\end{split}
\end{equation}
We can continue to iterate so to obtain components of the scalar that satisfy the wave equation at higher orders. The r-dependence of the next higher order piece is informed by the lower order piece before it.

The forms of equations \cref{scalarsol0} and \cref{scalarsol1} will be useful as we proceed in this paper to understand where the type-II double copy works and where it may not.  

\subsection{General Solution for the Wave Equation on the Fluids Background}
We begin with a look at a generic Maxwell spinor, $f_{AB}$, and examine the conditions necessary for it to satisfy Maxwell's equations.Note that in our conventions (see appendix \cref{NPappendix}) our principal null spinors are, 
\begin{equation}
	o_A = \frac{1}{\sqrt{2}}\{1,1\} \quad \textrm{and} \quad i_A = \frac{1}{\sqrt{2}}\{1,-1\}.
\end{equation} 
A general Maxwell spinor may thus have the form, 
\begin{equation}
	f_{AB} = P(\tau,r,z,\bar{z}) o_{(A}\iota_{B)} + Q(\tau,r,z,\bar{z}) o_{(A}o_{B)} + R(\tau,r,z,\bar{z}) \iota_{(A}\iota_{B)},  \label{pqr_ansatz}\\
\end{equation}
or equivalently (suppressing the functional dependence), 
\begin{equation}
	f_{AB} = \frac{1}{2}
	\begin{pmatrix}
		P + Q+R & Q-R\\
		Q-R & 	-P+Q+R
	\end{pmatrix},
\end{equation}
where, 
\begin{equation}
	\bar{f}_{AB} = (f_{AB})^* = (P\to\bar{P},\,\,Q\to\bar{Q}).
\end{equation}
The corresponding Maxwell tensor for this spinor has the form, 
\begin{equation}
	F_{ab} = \frac{1}{4}\begin{pmatrix}
		0 & -(P+\bar{P}) & -i(Q-\bar{Q}+ R - \bar{R}) & (Q+\bar{Q}- R - \bar{R})\\
		(P+\bar{P}) & 0 & -i(Q-\bar{Q}- R + \bar{R}) & (Q+\bar{Q}+ R + \bar{R})\\
		i(Q-\bar{Q}+ R - \bar{R}) & i(Q-\bar{Q}- R + \bar{R}) & 0 & i(P-\bar{P})\\
		-(Q+\bar{Q}- R - \bar{R}) & -(Q+\bar{Q}+ R + \bar{R}) & -i(P-\bar{P}) & 0
	\end{pmatrix}.
\end{equation}

Using the vierbiens and noting the frame tensor we can construct the Maxwell tensor as, $F_{\mu\nu}\equiv F_{ab}\,e_{\mu}^{\,\,a} e_{\nu}^{\,\,b}$, and then write the exterior derivatives $dF$ and $d^*F$,
\begin{align}
	\rm{dF}_{\tau rx}&=-\partial_x(P+\bar{P})+\frac{i}{\sqrt{\lambda}}\partial_{r}[r(R-\bar{R}) + (Q-\bar{Q})] + 2 i \sqrt{\lambda} \partial_{\tau}[R-\bar{R}], \label{dftrx}\\
	\rm{dF}_{\tau r y}&=-\partial_y(P+\bar{P})+ \frac{1}{\sqrt{\lambda}}\partial_{r}[r(R+\bar{R}) - (Q+\bar{Q})] + 2  \sqrt{\lambda} \partial_{\tau}[R+\bar{R}],\label{dftry}\\
	\rm{dF}_{\tau x y} &= i \partial_{\tau} (P-\bar{P}) + \frac{1}{\sqrt{\lambda}} \partial_{x}[r(R+\bar{R}) - (Q+\bar{Q})]
	-\frac{i}{\sqrt{\lambda}} \partial_{y}[r(R-\bar{R}) + (Q-\bar{Q})], \label{dftxy}\\
	\rm{dF}_{r x y}& = i \partial_{r} (P-\bar{P}) - 2 \sqrt{\lambda} \partial_x (R + \bar{R}) + 2 i \sqrt{\lambda} \partial_y (R - \bar{R}), \label{dfrxy}\\
	\rm{d^*F}_{\tau rx}&=i\partial_x(P-\bar{P})+\frac{1}{\sqrt{\lambda}}\partial_{r}[r(R+\bar{R}) + (Q+\bar{Q})] + 2 \sqrt{\lambda} \partial_{\tau}[R+\bar{R}],\label{dstarftrx} \\
	\rm{d^*F}_{\tau ry}&=i\partial_y(P-\bar{P})-\frac{i}{\sqrt{\lambda}}\partial_{r}[r(R-\bar{R}) - (Q-\bar{Q})] - 2 i \sqrt{\lambda} \partial_{\tau}[R-\bar{R}], \label{dstarftry}\\
	\rm{d^*F}_{ rxy}&=\partial_r (P + \bar{P}) + 2 \sqrt{\lambda} \partial_y (R + \bar{R}) + 2 i \sqrt{\lambda} \partial_x (R - \bar{R}),  \label{dstarfrxy} \\
	\rm{d^*F}_{\tau x y} &= \partial_{\tau} (P+\bar{P}) - \frac{i}{\sqrt{\lambda}} \partial_{x}[r(R-\bar{R}) - (Q-\bar{Q})]
	-\frac{1}{\sqrt{\lambda}} \partial_{y}[r(R+\bar{R}) + (Q+\bar{Q})].  \label{dstarftxy}
\end{align}
In order to correctly identify order by order solutions to the Maxwell's equations as written above we first identify expansions in functions P (similarly for Q and R) in $\lambda$, 

\begin{equation}\label{pexpansionnoi}
	\rm P \equiv P^{0} \lambda^{0} + P^{1/2} \lambda^{1/2} + P^{1} \lambda^{1} + O(\lambda^{3/2}).
\end{equation}
By utilizing the expansion above we see that the lowest order contribution fo Maxwell's equations is singularly provided by the $O(\lambda^{-1/2})$ coefficients in \eqref{dftrx} to \eqref{dstarftxy},
\begin{align}
	&\rm{dF}_{\tau rx}=\partial_{r}[r(R^{0}-\bar{R}^{0}) + (Q^{0}-\bar{Q}^{0})] = 0,\\
	&\rm{dF}_{\tau r y}=-\partial_{r}[r(R^{0}+\bar{R}^{0}) - (Q^{0}+\bar{Q}^{0})]= 0,\\
	&\rm{dF}_{r x y} = 0, \\
	&\rm{dF}_{\tau x y} =\partial_{x}[r(R^{0}+\bar{R}^{0}) - (Q^{0}+\bar{Q}^{0})] \nonumber \\&\hspace{3cm} - i \rm\partial_{y}[r(R^{0}-\bar{R}^{0}) + (Q^{0}-\bar{Q}^{0})], \\
	&\rm{d^*F}_{\tau rx}=\partial_{r}[r(R^{0}+\bar{R}^{0}) + (Q^{0}+\bar{Q}^{0})] = 0,\\
	&\rm{d^*F}_{\tau r y}=-\partial_{r}[r(R^{0}-\bar{R}^{0})- (Q^{0}-\bar{Q}^{0})]= 0,\\
	&\rm{d^*F}_{r x y} = 0, \\
	&\rm{d^*F}_{\tau x y} =\partial_{x}[r(R^{0}-\bar{R}^{0}) - (Q^{0}-\bar{Q}^{0})] \nonumber \\&\hspace{3cm} - \rm i \partial_{y}[r(R^{0}+\bar{R}^{0}) + (Q^{0}+\bar{Q}^{0})].
\end{align}
In the next subsection we will see how an identification of the Weyl scalars with these functions P, Q and R at different orders in $\lambda$ corresponds to solving Maxwell's equations.
\subsection{Mapping the Weyl Scalars to P,Q and R and Conditions On Solving Maxwell's Equations} 
A more correct approach to solving Maxwell's equations for the case at hand is to consider solutions order by order in the expansion parameter for the Maxwell spinor and map them to the Weyl Tensor. Note that we have the following non-vanishing orders of the Maxwell spinor. 

\begin{table}[H]
	\centering 
	\begin{tabular}{c|c|c}
		& Lowest Order Non-Vanishing Component&Error at $O(\lambda^k)$ \\ \hline
		$\Psi_{0}$ & First Non-Zero Term is an Error Term &{$O(\lambda)$} \\ \hline
		$\Psi_{1}$& {$O(\lambda^{1/2})$} &  {$O(\lambda^{3/2})$}\\ \hline
		$\Psi_{2}$& {$O(\lambda^{0})$} &{$O(\lambda^{1})$}\\ \hline
		$\Psi_{3}$&{$O(\lambda^{1/2})$}&{$O(\lambda^{3/2})$} \\ \hline
		$\Psi_{4}$ &{$O(\lambda^{0})$} &{$O(\lambda^{1})$}\\ \hline
	\end{tabular}     
\end{table}
Using the above, we map an order by order expansion in the Maxwell spinor functions P, Q and R onto the Weyl scalars. 

\begin{equation}\label{weylfull_psi}
	\begin{split}
		\frac{1}{S} f^{(1)}_{(AB}f^{(2)}_{CD)} =\Psi_0\, \iota_{A} \iota_{B} \iota_{C} \iota_{D} + 4\Psi_1\, \iota_{(A} \iota_{B} \iota_{C}o_{D)}& +  6 \Psi_2\, o_{(A} \iota_{B}o_{C} \iota_{D)}\\& + 4\Psi_3 \iota_{(A} o_{B} o_{C}o_{D)} +  \Psi_4\, o_{A} o_{B}o_{C}o_{D}.
	\end{split}
\end{equation}
Recall that we have identified what a general form of the Maxwell spinor may look like in terms of functions P, Q and R, thus for each of the spinors $f^{(1)}$ and $f^{(2)}$ we have, 
\begin{equation}
	f_{AB}^{(i)} = P(\tau,r,z,\bar{z})^{(i)} o_{(A}\iota_{B)} + Q(\tau,r,z,\bar{z})^{(i)} o_{(A}o_{B)} + R(\tau,r,z,\bar{z})^{(i)} \iota_{(A}\iota_{B)}.  \\
\end{equation}
Utilizing the above expansion for both Maxwell spinors along with the double copy, 
\begin{equation}\label{weylfull_f1f2}
	C_{ABCD} =\frac{1}{S_{II}} f^{(1)}_{(AB}f^{(2)}_{CD)},\\
\end{equation}
we now have, 
\begin{equation}
	\begin{split}\label{weylfull_pqr}
		C_{ABCD} =& \frac{1}{S_{II}} \biggr\{[ P^{(1)} P^{(2)} +(Q^{(1)} R^{(2)} + Q^{(2)} R^{(1)}) ] o_{(A} \iota_{B}o_{C} \iota_{D)} \\&+   [Q^{(1)} Q^{(2)}] o_{A} o_{B}o_{C}o_{D} + [ R^{(1)} R^{(2)}\iota_{A} \iota_{B} \iota_{C} \iota_{D} +  (P^{(1)} R^{(2)} + P^{(2)} R^{(1)})]  \iota_{(A} \iota_{B} \iota_{C}o_{D)} \\&+ [(Q^{(1)} P^{(2)} + P^{(2)}Q^{(1)})] \iota_{(A} o_{B} o_{C}o_{D)} \biggr\}.
	\end{split}
\end{equation}
Matching the above Weyl spinor to the Weyl scalars via the double copy \cref{weylfull_f1f2} we have, 
\begin{align}
	\Psi_0 = & \frac{1}{S_{II}} R^{(1)} R^{(2)}=0 + O(\lambda),\label{psi0eq} \\
	\Psi_1 = &  \frac{1}{S_{II}}  (P^{(1)} R^{(2)} + P^{(2)} R^{(1)})=\Psi_{1}^{1/2}\lambda^{1/2} + O(\lambda^{3/2}),\label{psi1eq}\\
	\Psi_2 = & \frac{1}{S_{II}}  [ P^{(1)} P^{(2)} + (Q^{(1)} R^{(2)} + Q^{(2)} R^{(1)})] = \Psi_{2}^{0}\lambda^{0} + O(\lambda^{1}),\label{psi2eq}\\ 
	\Psi_3 = & \frac{1}{S_{II}} (Q^{(1)} P^{(2)} + P^{(2)}Q^{(1)}) = \Psi_{3}^{1/2}\lambda^{1/2} + O(\lambda^{3/2}),\label{psi3eq}\\
	\Psi_4 = &  \frac{1}{S_{II}}  Q^{(1)} Q^{(2)} = \Psi_4^{0} \lambda^{0} + O(\lambda),\label{psi4eq}
\end{align}
where the superscripts not in parenthesis on the Weyl scalars indicate the non-zero orders of the Weyl scalars. Following \eqref{pexpansionnoi}, we write expansions for functions P (and similarly for Q and R), for the two Maxwell spinors, 
\begin{equation}\label{pexp}
	P^{(i)} \equiv P^{(i),0} \lambda^{0} + P^{(i),1/2} \lambda^{1/2} + P^{(i),1} \lambda^{1} + O(\lambda^{3/2}),
\end{equation}
the first superscript index $^{(i)}$ indicates which Maxwell spinor the functions P,Q or R belong to while the second superscript index (not in parenthesis) follows the $\lambda$ order in the expansion. We now have the necessary components necessary to map the expansions in functions P, Q and R order by order to the non-zero terms in various Weyl scalars as stated in \eqref{psi0eq} to \eqref{psi4eq}. At the lowest informative order in this map, i.e. at $O(\lambda^{0})$, we have,
\begin{align}
	R^{(1),0} R^{(2),0} &= 0, \\
	P^{(1),0} R^{(2),0} + P^{(2),0} R^{(1),0} &= 0,\\
	 P^{(1),0} P^{(2),0 } + (Q^{(1),0} R^{(2),0} + Q^{(2),0}R^{(1),0}) &= \Psi_{2}^{0},\\ 
	Q^{(1),0} P^{(2),0} + Q^{(2),0}P^{(1),0} &= 0,\,\,\, \\
	Q^{(1),0} Q^{(2),0} &= \Psi_4^{0}.
\end{align}
The above expressions result in two kinds of meaningfully distinct solutions, 
\begin{enumerate}
\item the first with no P in either $f^{(1)}$ or $f^{(2)}$ at lowest order, \begin{equation}\label{pqrsol_pzero}
	P^{(1),0} = P^{(2),0}=0 \qquad Q^{(2),0} = \frac{\Psi_4^{0}}{Q^{(1),0}} S_{II}^{0}\qquad R^{(1),0} = 0 \qquad R^{(2),0} = \frac{\Psi_2^{0}}{\,Q^{(1),0} } S_{II}^{0},
\end{equation}
\item the second with no R in either $f^{(1)}$ or $f^{(2)}$ at lowest order,
\begin{equation}\label{pqrsol_rzero}
	R^{(1),0} = R^{(2),0} = 0 \quad P^{(1),0} = i{Q^{(1),0}}\sqrt{\frac{\Psi_2^0}{\Psi_4^0}} \quad P^{(2),0} = -i\frac{\sqrt{\Psi_2^0\,\Psi_4^0}}{\,Q^{(1),0}} S_{II}^{0} \quad Q^{(2),0} = \frac{\Psi_4^{0}}{Q^{(1),0} } S_{II}^{0}.
\end{equation}
\end{enumerate}

In the subsequent subsections we examine Maxwell and scalar equations associated with the two solutions above in more detail. We will see that only the second of these solutions, \eqref{pqrsol_rzero},  results in a viable single copy, i.e. we satisfy both Maxwell's and scalar equations of motion upto $O(\lambda^{1/2})$, while the first solution \eqref{pqrsol_pzero} does not provide a non-vanishing scalar.

\subsubsection{Maxwell's Equations with Vanishing function P in both spinors}

Functions P, Q and R as identified by the first solution in \cref{pqrsol_pzero} result in the following pair of Maxwell spinors, 
\begin{equation}
	\begin{split}\label{fabpqrsol1}
		f_{AB}^{(1)} &= Q^{(1),0} o_{(A}o_{B)} + O(\lambda^{1/2}), \\
		f_{AB}^{(2)} &= \frac{\Psi_4^{0}}{Q^{(1),0}} S_{II}^{0} o_{(A}o_{B)} + \frac{\Psi_2^{0}}{\,Q^{(1),0} } S_{II}^{0} \iota_{(A}\iota_{B)}+ O(\lambda^{1/2}).  
	\end{split}
\end{equation}
Since Maxwell spinor $f_{AB}^{(1),0}$ (at its lowest relevant order) above only depends on $Q^{(1),0}$, requiring Maxwell's equations ${\rm dF^{(1)} = 0}$ and $\rm{d^*F^{(1)}=0}$ to hold at their lowest corresponding order $O(\lambda^{-1/2})$ results in the following conditions,
\begin{equation}
	\partial_{i} (Q^{(1),0}-\bar{Q}^{(1),0}) = 0 \quad\text{and}\quad \partial_{i} (Q^{(1),0}+\bar{Q}^{(1),0}) = 0 \quad \text{where}\,\,i\in \{r,x,y\}.
\end{equation}
The above expressions imply that the function $Q^{(1)}$ is independent of $r,x,y$, i.e. we have, 
\begin{equation}\label{q1rind}
	Q^{(1),0} = Q^{(1),0}(\tau).
\end{equation}
Similarly for the second Maxwell spinor $f_{AB}^{(2)}$ we find that the following conditions must hold in order for Maxwell's equations ${\rm dF^{(2)} = 0}$ and ${\rm d^*F^{(2)}=0}$ to hold,
\begin{equation}
	\begin{split}
		&\partial_{r}[r(R^{(2),0}-\bar{R}^{(2),0}) - (Q^{(2),0}-\bar{Q}^{(2),0})] = 0, \\
		&\partial_{r}[r(R^{(2),0}+\bar{R}^{(2),0}) - (Q^{(2),0}+\bar{Q}^{(2),0})] = 0, \\
		&\partial_x[r(R^{(2),0}+\bar{R}^{(2),0}) - (Q^{(2),0}+\bar{Q}^{(2),0})] - i \partial_y [r(R^{(2),0}-\bar{R}^{(2),0}) + (Q^{(2),0}-\bar{Q}^{(2),0})] =0,\\
		&\partial_x[r(R^{(2),0}-\bar{R}^{(2),0}) - (Q^{(2),0}-\bar{Q}^{(2),0})] - i \partial_y [r(R^{(2),0}+\bar{R}^{(2),0}) + (Q^{(2),0}+\bar{Q}^{(2),0})] =0.
	\end{split}
\end{equation}
The first two expressions above can be re-written as, 
\begin{equation}
	\partial_r(rR^{(2),0}- Q^{(2),0}) = 0 \quad \text{and} \quad c.c.
\end{equation}
Substituting for functions ${\rm R^{(2),0}}$ and ${\rm Q^{(2),0}}$ from \cref{pqrsol_pzero} above, we have, 
\begin{equation}
	\partial_r\biggr(r \frac{\Psi_2^{0}}{\,Q^{(1),0} } S_{II}^{0}- \frac{\Psi_4^{0}}{Q^{(1),0}} S_{II}^{0}\biggr) = 0.
\end{equation}
Using the fact that ${\rm Q^{(1),0}}$ is r independent from \cref{q1rind}, that the Weyl scalars have no r-dependence from \cref{psisol}, and that the r dependence in the zeroth copy scalar is logarithmic in r from \cref{scalarsol0} we have, 
\begin{equation}
	\begin{split}
		\rm \frac{\Psi_2^{0}}{Q^{(1),0} } \partial_r \big(r \, (S^{(0)}_a (\tau,x,y) + &\rm \log(r) S^{(0)}_b(\tau,x,y))\big) \\\qquad-  &\rm \frac{\Psi_4^{0}}{Q^{(1),0}} \partial_r (S^{(0)}_a (\tau,x,y) + \log(r) S^{(0)}_b(\tau,x,y)) = 0.
	\end{split}
\end{equation}
Since functions ${\rm S_a}$ and ${\rm S_b}$ are r independent, the above expression must be solved for coefficients of $\rm{r}$, ${\rm \log(r)}$ and the only solution that is permissible is if the scalar vanishes.

\subsubsection{Maxwell's Equations with Vanishing function R in both spinors}
\label{sec:typeiisolution}

Given that the first solution in \cref{pqrsol_pzero} did not provide a viable double copy, we now proceed to consider the second set of solutions as provided in \cref{pqrsol_rzero}, where at lowest order the function R vanishes, i.e. the coefficient of $\iota_{(A}\iota_{B)}$ is vanishing in both $f^{(1)}$ and $f^{(2)}$, this provides the following pair of Maxwell spinors, 
\begin{equation}
	\begin{split}\label{fabpqrsol2}
		f_{AB}^{(1)} &=iQ^{(1),0}\sqrt{\frac{\Psi_2^0}{\Psi_4^0}}o_{(A}\iota_{B)}  +  Q^{(1),0} o_{(A}o_{B)} + O(\lambda^{1/2}), \\
		f_{AB}^{(2)} &= -i\frac{\sqrt{\Psi_2^0\,\Psi_4^0}}{\,Q^{(1),0}} S_{II}^{0}o_{(A}\iota_{B)} + \frac{\Psi_4^{0}}{Q^{(1),0} } S_{II}^{0} o_{(A}o_{B)} + O(\lambda^{1/2}). 
	\end{split}
\end{equation}
The lowest order, $O(\lambda^{-1/2})$, Maxwell's equations when solved provide solutions to the  $O(\lambda^0)$ components for functions, i.e. $Q^{(1),0}$ and $Q^{(2),0}$, specifically we find the following conditions,
\begin{equation}
	\partial_{k} (Q^{(i),0}-\bar{Q}^{(i),0}) = 0 \,\,\text{and}\,\, \partial_{k} (Q^{(i),0}+\bar{Q}^{(i),0}) = 0 \,\, \text{where}\,\,k\in \{r,x,y\}\,\,\text{\&}\,\,i\in \{1,2\}.
\end{equation}
These differential equations imply that $Q^{(1),0}$ and $Q^{(2),0}$ are functionally independent of r, x and y and may only carry a dependence on Rindler time coordinate $\tau$, that is we have, 
\begin{equation}\label{qcond0}
	Q^{(i),0}(\tau,r,x,y) = Q^{(i),0}(\tau).
\end{equation}
Recall from \cref{pqrsol_rzero} we have for $Q^{(2),0}$ (i.e. coefficient of $o_{(A}o_{B)}$ in $f_{AB}^{(2),0}$, \cref{fabpqrsol2}), 
\begin{equation}\label{q2cond0}
\rm Q^{(2),0} = \frac{\Psi_4^{0}}{Q^{(1),0} } S_{II}^{0}.
\end{equation}
Consistency between \cref{qcond0} and \cref{q2cond0}, i.e. that both ${\rm Q^{(i),0}}$ are only time dependent  implies, 
\begin{equation}
	\rm \partial_{\,k}\big({\Psi_4^0 \,S_{II}^{0}}\big)= 0 \,\, \text{where}\,\,k\in \{r,x,y\},
\end{equation}
the r-independence of Weyl scalars \cref{psisol} will now  require that the zeroth copy scalar $\rm{S_{II}^{\,0}}$ must be r-independent and further its x,y dependence is inversely proportional to the Weyl scalar $\Psi_4$. That is we have,

\begin{equation}\label{scalarxydep}
\rm	\partial_{r} S_{II}^{\,\,0} = 0  \quad \text{and}\quad S_{II}^{\,\,0}(x,y) \sim \frac{1}{\Psi_4^{0}}.
\end{equation}
The above type-II scalar satisfies lowest order scalar equation of motion \cref{lambdaminus1S}. We thus have a lowest order type-II zeroth and single copies that satisfy the associated lowest order Maxwell and scalar equations of motion.

We can continue our iterative procedure, following \cref{lambdazeroS} we see that the next higher order component of the type-II scalar will be sourced via the Laplacian of ${\rm S_{II}^{\,\, 0}(\tau,x,y)}$. 

Similarly, examining Maxwell's equations at the next higher order allows us to put constraints on the functions $\rm P^{(i),0}$. From (\cref{dfrxy}) in $\rm dF_{rxy}$ in \cref{dfrxy} and $\rm d^*F_{rxy}$ in \cref{dstarfrxy}, the $O(\lambda^{0})$ components give us the following condition,
\begin{equation} \label{dfrxy0}
	\partial_r P^{(i),0}(\tau,r,x,y) = 0.
\end{equation}
The functions $\rm P^{(i)}$ in terms of Weyl scalars in \cref{fabpqrsol2} depend on the functions $Q^{(i)}$, $S_{II}$ and the Weyl scalars as,
\begin{equation}
	P^{(1)} = i{Q^{(1),0}}	\sqrt{\frac{\Psi_2^0}{\Psi_4^0}} \qquad P^{(2)} = -i\frac{\sqrt{\Psi_2^0\,\Psi_4^0}}{\,Q^{(1),0}} S_{II}^{\,\,0}. 
\end{equation} 
The r independence of the functions on the right sides of the above expressions allows us to satisfy \cref{dfrxy0}. \\

However, in order to fully solve the next order Maxell equation, we further need to solve $\rm dF_{\tau,r,x},dF_{\tau,r,y}$ and $\rm dF_{\tau,x,y}$ in  \cref{dftrx,dftry,dftxy} and the associated Hodge-duals in \cref{dstarftrx,dstarfrxy,dstarftry}. From these expressions, we note that $\tau$, x and y derivatives for $\rm P^{(i),0}$ depend on  $O(\lambda^{1/2})$ components of $Q$ and $R$. We now need to map these $O(\lambda^{1/2})$ components of functions Q and R on to Weyl scalars in \cref{psi0eq,psi1eq,psi2eq,psi3eq,psi4eq}. Using \cref{weylfull_psi,weylfull_f1f2,weylfull_pqr} we thus identify at $O(\lambda^{1/2})$,  
\begin{align}
	Q^{(2),1/2} &= -\frac {Q^{(1),1/2} Q^{(2),0}}{\psi_4^0} \label{Q2_aby2},\\ 
	P^{(1),1/2} &= -\psi_2^0 \psi_3^{1/2} + \psi_4^0 \psi_1^{1/2},\\ 
	P^{(2),1/2} &= \frac{Q^{(2),0}\psi_1^{1/2}}{\psi_2^0}, \\ R^{(2),1/2} &= 0 \label{R2_1by2},\\ 
	R^{(1),1/2} &= \frac{Q^{(1),1/2}\psi_2^0}{\psi_4^0} \label{R1_1by2}.
\end{align}

As we see, from \cref{Q2_aby2,R1_1by2,R2_1by2} we are able to identify the remaining functional dependencies for the function P in terms of $O(\lambda^0)$ components of weyl scalars $\Psi_2$, $\Psi_4$ as well as functions $Q^{(i),0}$ and $Q^{(i),1/2}$. For our tetrad choice we are thus able to obtain leading order solutions to Maxwell's and scalar wave equations, however for next higher orders a different tetrad choice is necessary. Beyond $O(\lambda^{1/2})$ we see that further higher order solutions are restricted by error pieces in the metric. 

Consequently we thus observe that for type-II fluid-dual space-times, Maxwell spinors of the form \cref{fabpqrsol1} do not provide viable double copy structures, while Maxwell spinors of the form \cref{fabpqrsol2} can be shown to solve background wave and scalar equations at leading order within the double copy paradigm.

\section{Conclusions}
The Weyl double copy picture has been applied to a large class of metrics. While the initial proposal was specific to type D spacetimes \cite{Luna:2018dpt}, twistor space formulations \cite{White:2020sfn,Chacon:2021wbr} have shown that how the Weyl double copy applies to algebraically general spacetimes, i.e. all the way upto Petrov type I metrics. 
	
Here we revisit the fluid-gravity duality in the context of the double copy paradigm \cite{keeler2020navierstokes}. While in \cite{keeler2020navierstokes}, we constrained ourselves to type D and type N algebraically special spacetimes, here we extend the double copy picture to include more general type II spacetimes. 

Our approach in this paper has been guided by the two dimensional fluid stream function, which we utilize to map geometric Weyl scalars onto specific fluid solutions. Further, the stream function provides a convenient tool that shows how the dynamics of the fluid are related to that of the spacetime. For instance in \cref{onvorticityeq}, we show clearly how the dynamics of Weyl scalar $\Psi_2$ are very similar to the dynamics of the fluid vorticity.

In line with the above emphasis on the stream function, we see that the simplest type II fluid that satisfies constraints \cref{cabcd} as well as the scalar and Maxwell equations of motion can be obtained by linearly combining stream functions associated with type D and type N fluids metrics. 
 
While considering more general solutions, we find that not all classes of  consistent factorization of type II fluid solutions \cref{cabcd} provide a non-trivial zeroth copy that satisfies the scalar equation of motion. Perturbatively, using our tetrad in \cref{sec:typeiisolution}, we see that non-trivial single and zeroth copy fields that satisfy algebraic constraints and the equations of motion to lowest order are only possible if the coefficient of the bispinor $\iota_{(A}\iota_{B)}$ vanishes. 

Our approach is inherently limited by the perturbative nature of the fluids metric we consider, \cite{Bredberg:2011jq}. To alleviate such perturbative limitations one may instead consider a more general fluids picture as proposed in \cite{Pinzani-Fokeeva:2014cka} and construct associated single and zeroth copy fields. Another interesting extension of our work will be possible if one were to consider relativistic fluids instead \cite{Compere:2012mt}. 

The classical double copy similarly provides an important and insightful tool for studying black hole horizons \cite{Chawla:2023bsu}. Similarly, the fluid-gravity paradigm in its various formulations \cite{Damour:1978cg,Damour:1979wya,Bhattacharyya:2008ji,Bhattacharyya:2008kq,Hubeny:2011hd,Bredberg:2010ky,Bredberg:2011jq,Bredberg:2011xw,Lysov:2011xx} allow for a deeper understanding of horizon dynamics and symmetries as fluids. Thus a broader examinations various fluid-dual geometries can thus provide physically insightful perspectives towards how asymptotic symmetries of such geometries map to their associated single copy factorizations and fluid-duals. 

\section{Acknowledgements}
N.M and C.K would like to thank Tucker Manton for discussions on the topic. C.K. is supported by the U.S. Department of Energy under grant number DE-SC0019470 and by the Heising-Simons Foundation “Observational Signatures of Quantum Gravity” collaboration grant 2021-2818. 
%


\section*{Appendix}
\appendix

\section{Newman-Penrose formalism}\label{NPappendix}

Similar to our approach in \cite{keeler2020navierstokes}, we once again follow the Newman-Penrose formalism \cite{Stephani:2003tm}. This formalism allows us to rewrite various geometric and gauge degrees of freedom as outer products of spinors. Central to the formalism is a rewriting of the metric tensor in terms of a set of null tetrads.  
\begin{equation}
	\label{tetradmetric}
	g_{\mu\nu} = -l_{(\mu}n_{\nu)} + m_{(\mu}\overline{m}_{\nu)}.
\end{equation}
As outlined in \cite{Cocke1989}, one utilizes above tetrad set $\{l,n,m,\bar{m}\}$ to obtain a set of spin coefficients which are then utilized to write various Weyl scalars, $\{\Psi_0,\Psi_1,\Psi_2,\Psi_3,\Psi_4,\}$,
\begin{equation}\label{NP}
	\begin{split}
		\Psi_{0}&=D\sigma-\delta\kappa-(\rho+\bar{\rho}+3\varepsilon+\bar{\varepsilon})\sigma+(\tau-\bar{\pi}+\bar{\alpha}+3\beta)\kappa \\
		\Psi_{1}&=D\beta-\delta\varepsilon-(\alpha+\pi)\sigma-(\bar{\rho}-\bar{\varepsilon})\beta+(\mu+\gamma)\kappa+(\bar{\alpha}-\bar{\pi})\varepsilon \\
		\Psi_{2}&=D\mu-\delta\pi+(\varepsilon+\bar{\varepsilon}-\bar{\rho})\mu+(\bar{\alpha}-\beta-\bar{\pi})\pi+\nu\kappa-\sigma\lambda-R/12 \\
		\Psi_{3}&=\bar{\delta}\gamma-\Delta\alpha+(\rho+\varepsilon)\nu-(\tau+\beta)\lambda+(\bar{\gamma}-\bar{\mu})\alpha+(\bar{\beta}-\bar{\tau})\gamma \\
		\Psi_{4}&=\bar{\delta}\nu-\Delta\lambda -(\mu+\bar{\mu}+3\gamma-\bar{\gamma})\lambda+(3\alpha+\bar{\beta}+\pi-\bar{\tau})\nu,
	\end{split}
\end{equation}
where the following are directional derivatives, 
\begin{equation}
	D=l^{\mu}\nabla_{\mu}, \ \ \ \ \Delta=n^{\mu}\nabla_{\mu}, \ \ \ \ \delta=m^{\mu}\nabla_{\mu}, \ \ \ \ \bar{\delta}=\bar{m}^{\mu}\nabla_{\mu}.
\end{equation}
In particular \ref{Cocke1989} allows an easier computation of spin coefficients $\sigma,\kappa, \rho,..etc.$ by using partial derivatives. 
Finally in terms of these Weyl scalars one can rewrite the Weyl Spinor,
\begin{equation}
	C_{ABCD}=\Psi_0 \iota_{A}\iota_B\iota_C\iota_D-4\Psi_1 o_{(A}\iota_B\iota_C\iota_{D)}+6\Psi_2 o_{(A}o_B\iota_C\iota_{D)}-4\Psi_3 o_{(A}o_Bo_C\iota_{D)}+\Psi_4 o_Ao_Bo_Co_D.\\
\end{equation}
The spinors ${o_A,\iota_A}$ are related to the frame metric choice one makes. We will make explicit this connection now. The metric written in terms of vierbiens has the form,
\begin{equation}
	g_{\mu\nu} = e_{\mu}^{\,\,a} e_{\nu}^{\,\,b} \eta_{ab} \,\,\,\, \text{where}\,\,\,\,\eta_{ab} = \text{diag} \{-1,1,1,1\}.
\end{equation}
The frame metric $\eta_{ab}$ can itself be written as outer products of a tetrad set, this will allow us to make identifications between the vierbiens and the tetrad set.
\begin{equation}
	\begin{split}
		\eta_{ab} &= -\hat{l}_{(a}\hat{n}_{b)} + \hat{m}_{(a}\hat{\overline{m}}_{b)}\\
		\implies g_{\mu\nu} &=  e_{\mu}^{\,\,a} e_{\nu}^{\,\,b} (-\hat{l}_{(a}\hat{n}_{b)} + \hat{m}_{(a}\hat{\overline{m}}_{b)})\\
		\implies g_{\mu\nu} &= -l_{(\mu}n_{\nu)} + m_{(\mu}\overline{m}_{\nu)}
	\end{split}    
\end{equation}
Where in the last step we have made the identifications, 
\begin{equation}
	\label{vtot}
	e_{\mu}^{\,\,a}\hat{l}_a = l_{\mu} \hspace{0.3cm} e_{\mu}^{\,\,a}\hat{n}_a = n_{\mu} \hspace{0.3cm} e_{\mu}^{\,\,a}\hat{m}_a = m_{\mu} \hspace{0.3cm} e_{\mu}^{\,\,a}\hat{\overline{m}}_a = \overline{m}_{\mu}. 
\end{equation}
Now the tetrad set that reproduces the Minkowski frame metric is, 
\begin{equation}
	\begin{split}
		\hat{l}_a &= \frac{1}{\sqrt{2}}\{1,-1,0,0\}\\
		\hat{n}_a &= \frac{1}{\sqrt{2}}\{1,1,0,0\}\\
		\hat{m}_a &= \frac{1}{\sqrt{2}}\{0,0,i,1\}\\
		\hat{\overline{m}}_a &= \frac{1}{\sqrt{2}}\{0,0,-i,1\}.\\    
	\end{split}    
\end{equation}
The expression \cref{vtot} can be inverted to go from tetrads to vierbiens via the following, 

\begin{equation}\label{tetradtovierbien}
	\begin{split}
		&e_{\mu}^{\,\,0} = \frac{1}{\sqrt{2}}(l_{\mu}+n_{\mu}) \qquad \,\,\,\,\, e_{\mu}^{\,\,1} = \frac{1}{\sqrt{2}}(l_{\mu}-n_{\mu})\\
		&e_{\mu}^{\,\,2} = \frac{i}{\sqrt{2} }  (\bar{m}_{\mu}-m_{\mu}) \qquad
		e_{\mu}^{\,\,3} =  \frac{1}{\sqrt{2} }  (m_{\mu}+\bar{m}_{\mu}).
	\end{split}
\end{equation}

In order to obtain the spinors $\{o_A,\iota_A\}$ we write these {\bf{ISO(3,1)}} four vectors in an {\bf{SL(2,C)}} representation by contracting them with Pauli vectors, which in our conventions are the set, $\sigma^a_{A\dot{A}}=\{\mathbb{I},\vec{\sigma}\}$, 
\begin{equation}
	\begin{split}
		o_A o_{\dot{A}} &\equiv \hat{l}_a\, \sigma ^a_{A\dot{A}} = \frac{1}{2}  \begin{pmatrix}
			1 & 1\\
			1 & 1 \\ \end{pmatrix}\\
		&\implies o_A = \frac{1}{\sqrt{2}}\{1,1\}\\
		\iota_A \iota_{\dot{A}} &\equiv \hat{n}_a\, \sigma ^a_{A\dot{A}} = \frac{1}{2}  \begin{pmatrix}
			1 & -1\\
			-1 & 1 \\ \end{pmatrix}\\
		&\implies i_A = \frac{1}{\sqrt{2}}\{1,-1\}.\\  
	\end{split}
\end{equation}

Further noting that one can transform from {\bf SL(2,C)} left to right by complex conjugation we use the convention, 
\begin{equation}
	(o_A)^* \equiv o_{\dot{A}}
\end{equation}
This further verifies that the two remaining contractions will hold the following relations correctly (albeit simpler given that our choice of dyads is real), 
\begin{equation}
	\begin{split}
		\hat{m}_a\, \sigma ^a_{A\dot{A}} &= \frac{1}{2}  \begin{pmatrix}
			1 & 1\\
			-1 & -1 \\ \end{pmatrix} =    \iota_A o_{\dot{A}}\\
		\hat{\overline{m}}_a\, \sigma ^a_{A\dot{A}} &= \frac{1}{2}  \begin{pmatrix}
			1 & -1\\
			1 & -1 \\ \end{pmatrix} =   o_A \iota_{\dot{A}} \\
	\end{split}
\end{equation}
In the near horizon($\lambda$) expansion the vierbiens $e_{\mu}^{(0),a}$, as obtained by setting the velocities and pressures to zero in the fluids metric in equation \eqref{lambdaexpansion}) are, 
\begin{equation}
	\begin{split}\label{bkgvierbiens}
		e^{(0),0}_{\,\mu} &= \{\frac{1+r}{2 \sqrt{\lambda}},-\sqrt{\lambda},0,0\},\\
		e^{(0),1}_{\,\mu} &= \{\frac{1-r}{2 \sqrt{\lambda}},\sqrt{\lambda},0,0\},\\				e^{(0),2}_{\,\mu}&=\{0,0,1,0\},\\	
		e^{(0),3}_{\,\mu}&=\{0,0,0,1\}, 
	\end{split}
\end{equation}
which reproduce the background metric, 
\begin{equation}
	\begin{split}
		\label{rindlerlambda}
		\eta_{ab} e^{(0),a}_{\mu}e^{(0),b}_{\mu} &= g_{\mu\nu}^{0}
		=\begin{pmatrix}
			-\frac{r}{\lambda} & 1 & 0 & 0\\
			1 & 0 & 0 & 0\\
			0 & 0 & 1 & 0\\								0 & 0 & 0 & 1			
		\end{pmatrix}.
	\end{split}
\end{equation}
Similarly, we can construct the Maxwell tensor together, i.e. $F_{\mu\nu}=F_{ab}e^{(0),a}_{\mu}e^{(0),b}_{\nu}$.

The algebraic speciality of the spacetime is identified by the computation of various covariant combinations of the Weyl scalars, \
\begin{equation}\label{Weylcovariants}
	\begin{split}
		I&\equiv\Psi_0\Psi_4-4\Psi_{1}\Psi_{3}+3\Psi_{2}^{2}, \\
		J&\equiv \begin{vmatrix} 
			\Psi_{4}&\Psi_{3}&\Psi_{2}\\
			\Psi_{3}&\Psi_2 &\Psi_1 \\
			\Psi_{2}&\Psi_{1} &\Psi_{0}
		\end{vmatrix} ,\\
		K&\equiv \Psi_{1}\Psi_{4}^{2}-3\Psi_{4}\Psi_{3}\Psi_{2}+2\Psi_{3}^{3}, \\
		L&\equiv \Psi_{2}\Psi_{4}-\Psi_{3}^{2}, \\
		N&\equiv 12L^{2}-\Psi_{4}^{2}I.
	\end{split}
\end{equation}

Of interest to us are spacetimes referred to as being algebraically special. In order for the spacetime to be algebraically special, it must minimally satisfy \cite{Stephani:2003tm},
\begin{equation}\label{specialmetrics}
	I^3 - 27 J^2 = 0.
\end{equation}
Referred to as Petrov types, when algebraically special, such spacetimes may be most generically Petrov type-II spacetimes. With additional constraints they may further be subclassified into to be type D or type N spacetimes, as was the case in our previous paper \cite{keeler2020navierstokes}, or be further simplified to be type-III or type O spacetimes. 

\section{A Note on the Vorticity Equation} \label{onvorticityeq}

Using the Navier-Stokes equation it is possible to identify an expression for the vorticity of the fluid, recall the NS equation reads, 

\begin{equation}
\partial_{\tau} \,v_i - \eta \partial^2 v_i + \partial_i P + v_j \partial_j v_i = 0 \qquad \forall \quad i \in \{1,2\}
\end{equation}
Taking the curl of the Navier-Stokes equation we get, 
\begin{equation}
	\partial_{\tau} \epsilon_{ij} \partial_{i} v_j - \eta \partial^2 \epsilon_{ij} \partial_i v_j + v_k \partial_k \epsilon_{ij} \partial_i v_j = 0, 
\end{equation}
where the gradient of the pressure  vanishes following the curl. In terms of the vorticity, $\omega = \epsilon_{ij}v_iv_k$, the above expression has the form, 
\begin{equation}\label{NSVorticityEQ}
	\partial_{\tau} \omega - \eta \partial^2 \omega + v_k \partial_k \omega = 0. 
\end{equation}

For our purposes we would like to write this equation in terms of the stream function $\chi$ which provides the velocity of the fields as, 
\begin{equation}\label{streamtovelocity}
	v_i  = \epsilon_{ij} \partial_j \chi.
\end{equation}
As stated in \cref{psi2vorticity}, the vorticity of the fluid can thus be identified in terms of the stream function, 
\begin{equation}\label{psi2vorticityAppendix}
	\partial^2 \chi = -\omega(x,y).
\end{equation}
The above equation allows us to rewrite \cref{NSVorticityEQ} in terms of the stream function as, 
\begin{equation}
	\label{NSchi}
	i \partial_{\tau} \partial_{\bar{z}}\partial_z \chi - 4 i \eta \partial_z^2\partial_{\bar{z}}^2 \chi + 2 \partial_{\bar{z}}\chi \partial_{z} (\partial_{z}\partial_{\bar{z}}\chi ) - 2 \partial_z \chi \partial_{\bar{z}} (\partial_{z}\partial_{\bar{z}}\chi ) = 0. 
\end{equation}

Recall from \label{psisol} that we were able to rewrite Weyl scalars in terms of the stream function, 
\begin{equation}\label{psisolAppendix}
	\Psi_2 = i \partial_z \partial_{\bar{z}} \chi \qquad \Psi_4 = 2 i  \partial_{\bar{z}}^2 \chi.
\end{equation}
 
We can now identify Weyl scalar $\Psi_2$ (or $\Psi_4$) in \cref{NSchi}, finally allowing us to rewrite the fluid vorticity equation in \cref{NSVorticityEQ} as,
\begin{equation}
	\label{NSchiToPsi2}
	\partial_{\tau} \Psi_2 - 4 \eta \partial_z\partial_{\bar{z}} \Psi_2 - 2 i  (\partial_{\bar{z}}\chi \partial_{z}\Psi_2 - 2 \partial_z \chi \partial_{\bar{z}} \Psi_2 ) = 0.
\end{equation} 

Rearranging terms in \cref{NSchiToPsi2} and utilizing \cref{streamtovelocity}, we are able to express dynamics of weyl scalar $\Psi_2$ in a very similar form to that of the vorticity of the fluid in \cref{NSVorticityEQ},
\begin{equation}
	\label{NSchi}
	\partial_{\tau} \Psi_2 - 4 \eta \partial_z\partial_{\bar{z}} \Psi_2 +  v_z \partial_{z}\Psi_2 + v_{\bar{z}} \partial_{\bar{z}} \Psi_2  = 0. 
\end{equation} 
The above expression can be further canonicalized using a convective derivative, $D_{\tau}\equiv \partial_{\tau}+v_j \partial_j, \,\,\text{where}\,\, j \in \{z,\bar{z}\}$,

\begin{equation}\label{NSPsi2Canonical}
	4 \eta \partial_z\partial_{\bar{z}} \Psi_2 =  D_{\tau} \Psi_2 \qquad D_{\tau}\Psi_2 \equiv \frac{d\,\Psi_2(\tau,x,y)}{d\tau} =\partial_{\tau} \Psi_2 + v_j \partial_j \Psi_2.  
\end{equation}

Note that the Navier-Stokes equation can also be written more compactly in terms of a convective derivative, $D_{\tau}$, 
\begin{equation}\label{NSChiCanonical}
	\eta \partial^2 \omega =  D_{\tau} \omega \qquad D_{\tau}\omega \equiv \frac{d\,\omega(\tau,x,y)}{d\tau} =\partial_{\tau} \omega + v_j \partial_j \omega.  
\end{equation}

Two simple classes of solutions that emerge involve studying the vorticity equation in the inviscid regime (with the viscosity turned off) or by turning off the directional derivative, which reduces to the heat equation, that is we have, 

\begin{enumerate}
	\item Setting the viscosity off. This gives $D_{\tau}\omega=0$, these solutions are often referred to as point vortex solutions. Following a similar procedure while searching for a suitable Weyl scalar we have, 
	\begin{equation}\label{NSheateqsplit}
		D_{\tau}\Psi_2=0.
	\end{equation}
	\item Fluid Solutions where $v_j \partial_j \omega = 0$, these reduce the vorticity equation to the heat equation, and such solutions are referred to as Lamb-Oseen vortices. In terms of the Weyl scalar this corresponds to, 
	\begin{equation}\label{NSLambOseenSplit}
		4 \eta \partial_z\partial_{\bar{z}} \Psi_2 =  \partial_{\tau} \Psi_2.
	\end{equation}
\end{enumerate}
The above equations: \cref{NSchi} to \cref{NSLambOseenSplit} can thus be utilized to construct additional fluid solutions as necessary. 
\bibliographystyle{hunsrt.bst}
{\small
	\bibliography{helmholtz}}
\end{document}